\documentclass[10pt, conference]{IEEEtran}
\IEEEoverridecommandlockouts
\usepackage[
  backend=biber,
  maxcitenames=3,
  minbibnames=1,
  mincitenames=1,
  maxbibnames=3,
  uniquelist=false,
  style=ieee,
  url=false,
  eprint=false,
  isbn=false,
]{biblatex}
\usepackage{graphicx}
\usepackage{amsmath}
\usepackage{amssymb}
\usepackage[colorlinks=true,urlcolor=teal,
            linkcolor=teal,citecolor=teal]{hyperref}
\usepackage{multirow}
\usepackage{csquotes}
\usepackage{xcolor}
\usepackage{tikz}
\usepackage{yquant}
\usepackage{braket}
\usepackage{subcaption}
\usepackage[T1]{fontenc}
\usepackage{censor}
\usepackage{xspace}
\usepackage{tabularx}
\usepackage{booktabs}

\usepackage{tikz}
\usetikzlibrary{arrows.meta,positioning, calc, shadows.blur}

\AtBeginBibliography{\footnotesize}
\newboolean{anonymous}
\setboolean{anonymous}{false}
\newcommand{\eg}{\emph{e.g.}\xspace}
\newcommand{\ie}{\emph{i.e.}\xspace}
\newcommand{\etal}{\emph{et al.}\xspace}

\ifbool{anonymous}{}{}
\ifbool{anonymous}{}{\renewcommand{\blackout}[1]{#1}}
\ifbool{anonymous}{\newcommand{\genemail}[2]{\href{mailto:xxx.xxx@xx.xx}{\blackout{#2}}}}{\newcommand{\genemail}[2]{\href{#1}{#2}}}

\newcommand{\repro}{\url{https://github.com/lfd/make_some_noise}}
\ifbool{anonymous}{\renewcommand{\repro}{\href{https://thiscatexists.com/}{\blackout{https://github.com/lfd/make\_some\_noise}}}}{}

\makeatletter
\DeclareRobustCommand\rvdots{%
  \vbox{%
    \baselineskip4\p@\lineskiplimit\z@%
    \kern-\p@%
    \hbox{.}\hbox{.}\hbox{.}%
  }%
}

\definecolor{lfd1}{HTML}{000000} % For background use, white is colour 1
\definecolor{lfd2}{HTML}{E69F00}
\definecolor{lfd3}{HTML}{999999}
\definecolor{lfd4}{HTML}{009371}
\definecolor{lfd5}{HTML}{BEAED4}
\definecolor{lfd6}{HTML}{ED665A}
\definecolor{lfd7}{HTML}{1F78B4}
\definecolor{lfd8}{HTML}{009371}

\begin{document}

\title{Make Some Noise! Measuring Noise Model Quality in Real-World Quantum Software}

\author{%
\IEEEauthorblockN{\blackout{Stefan Raimund Maschek, Jürgen Schwitalla}}
\IEEEauthorblockA{\blackout{\textit{science + computing AG, Eviden}}\\
    \blackout{Tübingen, Germany}\\
    \{\genemail{mailto:stefan-raimund.maschek@eviden.com}{stefan-raimund.maschek}, \genemail{mailto:juergen.schwitalla@eviden.com}{juergen.schwitalla}\}@\blackout{eviden.com}}
\and
\IEEEauthorblockN{\blackout{Maja Franz}}
\IEEEauthorblockA{\blackout{\textit{Technical University of}}\\
    \blackout{\textit{Applied Sciences Regensburg}} \\
    \blackout{Regensburg, Germany} \\
    \genemail{mailto:maja.franz@othr.de}{maja.franz@othr.de}}
    \and 
\IEEEauthorblockN{\blackout{Wolfgang Mauerer}}
\IEEEauthorblockA{\blackout{\textit{Technical University of}}\\
    \blackout{\textit{Applied Sciences Regensburg}}\\
    \blackout{\textit{Siemens AG, Technology}}\\
    \blackout{Regensburg/Munich, Germany}\\
    \genemail{mailto:wolfgang.mauerer@othr.de}{wolfgang.mauerer@othr.de}}
}

\maketitle

\begin{abstract}
  Noise and imperfections are among the prevalent challenges in quantum software engineering for current NISQ systems. They will remain important in the post-NISQ area, as logical, error-corrected qubits will be based on software mechanisms.
  As real quantum hardware is still limited in size and accessibility,
  noise models for classical simulation---that in some cases can exceed dimensions of actual 
  systems---play a critical role in obtaining insights into quantum algorithm performance, and the properties of mechanisms for error correction and mitigation.

  We present, implement and validate a tunable noise model building on the Kraus channel formalism on a large scale quantum simulator system (\emph{Qaptiva}).
  We use empirical noise measurements from IBM quantum (IBMQ) systems to calibrate the model and create a realistic simulation environment.
  Experimental evaluation of our approach with Greenberger-\hspace*{0mm}Horne-\hspace*{0mm}Zeilinger (GHZ) state preparation and QAOA applied to an
  industrial use-case validate our approach, and demonstrate 
  accurate simulation of hardware behaviour at reasonable computational cost.

  We devise and utilise a method that allows for determining the quality of noise models for larger problem instances than is possible with existing metrics in the literature. To identify potentials of future quantum software and algorithms, we extrapolate the noise model to future partially fault-tolerant systems, and give insights into the interplay between hardware-specific noise modelling and hardware-aware algorithm development.
\end{abstract}


\IEEEpeerreviewmaketitle

\section{Introduction}
\label{sec:intro}

Despite recent advancements in constructing error-corrected and fault-tolerant quantum computers~\cite{google_willow24, morteza25}, noise and imperfections remain dominant challenges in current noisy intermediate-scale quantum (NISQ) systems~\cite{bharti22}.
Therefore, there persists a need to acknowledge noisy hardware and in particular noisy models of quantum hardware when engineering quantum software and algorithms.
For the path toward achieving practical quantum advantage, it is essential to understand which types of noise impact specific computations and to what extent.
A deep understanding of hardware limitations can guide the development and deployment of hardware-specific quantum algorithms, fostering a hardware-software co-design paradigm~\cite{safi23, wintersperger22}.
A critical component in this approach is the ability to accurately simulate the behaviour of current quantum processing units (QPUs) using noise models.

Additional advantages of a noisy simulation over an execution on real QPUs include the ability to analyse quantum algorithms at any intermediate point in a circuit and the ability to directly obtain probabilities (and even amplitudes) instead of relative frequencies. 
Furthermore, a noise model also represents a fictitious parametrisable QPU that can be used as a test bed for future quantum algorithms.

While the theoretical modelling of various noise sources is well-established~\cite{nielsen10}, simulating the exact behaviour of a specific physical QPU remains challenging.
However, it is possible to develop increasingly accurate noise models that capture the nuances of real-world quantum hardware.
The fidelity of these models heavily depends on the availability and accuracy of detailed noise data provided by hardware vendors, which is not always fully accessible.

Determining the quality of a noise model is essential for ensuring that the theoretical predictions align with experimental results.
As a crucial step in software engineering, it allows developers to identify and correct inaccuracies in their models, ultimately leading to more robust and maintainable quantum software.
Without this step, the gap between simulation-based designs and real-world quantum computations could widen, potentially delaying the realisation of practical quantum algorithms.
However, typical similarity measures between simulation and experiment, such as the Hellinger distance (HD)~\cite{Georgopoulos_2021}, quickly reach its limits when increasing the number of qubits due to a lack of sufficient statistics.
Therefore, we present a metric that intuitively quantifies the quality of a noise model, based on the overall quality of the corresponding practical quantum algorithms, to which the model is applied to.

In this work, we furthermore evaluate the accuracy of IBM's publicly available noise data~\cite{ibm} using \emph{Qaptiva}'s noise model implementation~\cite{Qaptiva}.
Our evaluation comprises experimental studies using proof-of-concept circuits, such as the Greenberger-Horne-Zeilinger (GHZ) circuit~\cite{greenberger90}.
Additionally, we demonstrate the practical relevance of our approach by applying it to a real-world problem:
implementing the quantum approximate optimisation algorithm (QAOA)~\cite{farhi14} to address job shop scheduling problems~\cite{xiong22}.

Furthermore, we extrapolate the performance of quantum algorithms to future QPUs by calibrating our noise models to align with more fault-tolerant systems, which are not currently available.
This allows the development and evaluation of hardware-aware and fault-tolerant algorithms, which, in their current state, cannot be implemented on actual quantum hardware.

Our software stack that integrates the noise model delivers the toolset to build a realistic and hardware-accurate simulation environment. To integrate it with the quantum software ecosystem, we utilise the \emph{QUARK} framework~\cite{rudi22}.

The remainder of this article is structured as follows:
\autoref{sec:se_perspective} gives an overview of the topic in the domain of quantum software engineering, reviews related work and introduces the theoretical background to noise modelling.
In \autoref{sec:methodology}, we describe our methodology, including our parametrised noisy simulator, as well as the utilised benchmark circuits and metrics.
\autoref{sec:exp_results} explains our experimental results, which are extrapolated to future QPUs in \autoref{sec:scaled_fid} and discussed in \autoref{sec:disc}.
Finally, we conclude in \autoref{sec:concl}.

\section{Noise Modelling: Software Perspective}
\label{sec:se_perspective}
As long as imperfections in quantum hardware persist, the development of quantum software and algorithms must account for noisy hardware, which is built on diverse paradigms and exhibits distinct properties~\cite{greiwe23}.
These unique characteristics significantly influence the types of noise that can occur on a given platform.
Consequently, understanding these noise patterns is essential when designing quantum algorithms or broader quantum software solutions.

However, with limited and costly access to real QPUs, the next best alternative is to develop accurate noise models of these systems.
These models serve as critical tools for simulating and predicting the behaviour of quantum hardware in various scenarios, albeit simulations ---especially in the noisy case--- are only feasible for limited system sizes.

\subsection{Noisy Simulation Challenges}
From a quantum software engineering perspective, many existing approaches focus on higher-level abstractions rather than low-level noise modelling.
This abstraction, while useful for theoretical advancements, risks oversimplifying the complexities of noise in practical systems.
A general noise model may fail to capture the unique characteristics of a specific QPU, necessitating the development of detailed noise models tailored to individual hardware platforms.

The accuracy of these detailed noise models depends on two critical factors:
(1) the granularity and detail of the modelling process and (2) the availability and precision of \emph{noise data}, which describes the real-world behaviour of the QPU.
Without reliable noise data, even the most sophisticated modelling techniques may fall short of capturing the true dynamics of a quantum system.

Moreover, when treating a noise model as a black box, important insights into the behaviours of near-term quantum algorithms and systems may be lost.
This abstraction can hinder the ability to fine-tune algorithms or hardware configurations to mitigate specific types of noise, ultimately limiting the practical advancements in quantum computing. 

\subsection{Related Work}
The integration of noise models into the quantum software stack, as a drop-in replacement of real QPUs for test and evaluation purposes of quantum algorithms, has been explored in previous studies~\cite{becker22, resch21, tomesh22}.
Additionally, a growing body of literature examines the performance of quantum algorithms---such as certain state-preparation tasks~\cite{leymann20}, or variational algorithms including QAOA~\cite{thelen24, safi23, wintersperger22} and quantum machine learning~\cite{franz24}---under the influence of various noise conditions.

Given that variational algorithms are considered the most promising for near-term quantum systems~\cite{bharti22}, insights from these studies are particularly valuable for quantum software engineering, as highlighted by Greiwe~\etal~\cite{greiwe23}, who demonstrate general noise modelling tailored to NISQ algorithms.
However, a noticeable research gap exists in directly comparing noise models with real hardware.

While comprehensive benchmarks of actual quantum hardware have been conducted, for example by Li~\etal~\cite{li23}, most noise model evaluations itself are rather sparse, and focus on conceptual, instead of practical quantum algorithms.
For instance, Bravo-Montes~\etal~\cite{bravo_montes24} evaluated various noise models and software simulators against QPUs for small-scale logical, arithmetical, and error-correction circuits, and Grover's algorithm.
Similarly, Georgopoulos~\etal~\cite{Georgopoulos_2021} assessed different noise modelling techniques in quantum walk contexts.

These studies typically measure noise model effectiveness using state fidelities between simulations and experimental results.
A notable exception is the work by Weber~\etal~\cite{weber24}, who introduced algorithm-specific quality metrics for variational algorithms.
Our work explores a similar direction by introducing refined metrics tailored to specific applications and offering guidance on incorporating insights from noise model performance assessments into the quantum software stack.

\subsection{Theoretical Background}

The basis for the noise model is described in Refs.~\cite{Georgopoulos_2021,Krantz_2019} and is implemented using the Qaptiva libraries~\cite{Qaptiva}.
The following description makes use of the Kraus formalism with each Kraus operator $K$ acting on the density matrix $\rho$ as
\begin{align}
    \rho \rightarrow \sum_i K_i \rho K_i^\dagger.
\end{align}
The noise model is accounting for three main noise channel categories,
(1) amplitude damping and dephasing (\ie environmental effects, decoherence)
(2) depolarisation (gate errors) and
(3) state preparation and measurement (SPAM) errors.

\textbf{Amplitude Damping} describes the natural decay of the excited state $\ket{1}$ to the ground state $\ket{0}$ due to energy exchange with the environment.
The corresponding Kraus operators are
\begin{align}
    E_0 (t) & = \ket{0} \bra{0} + \sqrt{\exp(-t/T_1)}\ket{1} \bra{1}, \\
    E_1 (t) & = \sqrt{1-\exp(-t/T_1)} \ket{0} \bra{1},
\end{align}
parametrised by the relaxation time $T_1$.

\textbf{Pure Dephasing} similarly describes the transition of a quantum system towards classical behaviour, defined by the Kraus operators
\begin{align}
    E_0 (t) & = \ket{0} \bra{0} + \sqrt{1-p(t)}\ket{1} \bra{1}, \\
    E_1 (t) & = \sqrt{p(t)}\ket{1} \bra{1},
\end{align}
with the time-dependent probability,
\begin{align}
    p(t) = 1-\exp(-\frac{2t}{T_\varphi}) \\
    \text{where}\qquad {T_\varphi} = \left(\frac{1}{T_2} - \frac{1}{T_1}\right)^{-1}, \label{eq:tphi}
\end{align}
with the pure dephasing time $T_\varphi$.

The combined effect of amplitude damping and dephasing results in the following evolution of the one qubit density matrix, which is the desired phenomenological description:
\begin{align}
  \label{eq:dephasing}
\left[ {\begin{array}{cc}
a & b \\
b^* & 1-a \\
\end{array} } \right]
  \rightarrow
\left[ {\begin{array}{cc}
(a-1)e^{-\frac{t}{T_1}} + 1 & b e^{-\frac{t}{T_2}}\\
b^*e^{-\frac{t}{T_2}} & (1-a) e^{-\frac{t}{T_1}}\\
\end{array} } \right]
\end{align}

In applying these channels we apply decoherence during idle times, which is a slightly different approach as in Ref.~\cite{Georgopoulos_2021}.
Our modelling is motivated by the assumption that the qubits interact with the environment especially during idle times of the qubits.

\textbf{Depolarisation} channels are parametrised in this model by the one- and two-qubit gate fidelities $\mathcal{F}_{\text{1-gate}}$ and $\mathcal{F}_{\text{2-gate}}$.
The depolarisation noise quantum channel for one-qubit gates is defined by the Kraus operators
\begin{align}
    K_0 = \sqrt{1-p} \, I, &  &
    K_i = \sqrt{p/3} \, \sigma_i, \qquad \text{for}\,  i = 1,2,3,
\end{align}
with $\sigma_i$ the Pauli matrices, 
and for two-qubit gates by
\begin{align}
    \{K_i \otimes K_j \}_{i,j=0\ldots3}.
\end{align}
The depolarisation noise quantum channel is applied after the (noiseless) quantum gate.

\textbf{State preparation and measurement (SPAM) errors} can be modelled with the following Kraus channels.
The measurement error is described by
\begin{align}
    E_1 &= \varepsilon^\text{meas}_0 \ket{0}\bra{0} + (1-\varepsilon^\text{meas}_1) \ket{1} \bra{1}, \\
    E_0 &= (1-\varepsilon^\text{meas}_0) \ket{0}\bra{0} + \varepsilon^\text{meas}_1 \ket{1} \bra{1},
\end{align}
which is applied immediately before a measurement, with the error rates $\varepsilon^\text{meas}_0$/$\varepsilon^\text{meas}_1$ for mismeasuring  $\ket{0}$/$\ket{1}$ respectively.
The initialisation error, applied as the first operation, is described by the Kraus operators
\begin{align}
    E_0^\text{init} = (1-\varepsilon^\text{init}_0) \ket{0}\bra{0} + \varepsilon^\text{init}_1 \ket{1} \bra{1}.
\end{align}

The noise parameters are summarised in~\autoref{tab:noise_parameter}.
These parameters are experimentally measurable and generally vary among the individual qubits/gates of the physical hardware.
The QPU manufacturer provides these parameters in the course of the hardware calibration along with the corresponding uncertainties.

\begin{table}[h]
  \begin{tabularx}{\linewidth}{l|p{2.5cm}Xp{2.5cm}}
                                           & \textbf{Parameter}           & \textbf{Ref. Values}   & \textbf{Noise Channel}                   \\
        \toprule
        $T_1$                            & Relaxation Time               & $10^{-5}~s$      & Amplitude Damping               \\
        $T_2$                            & Dephasing Time      & $\leq 2T_1$      & \multirow{3}{*}{Dephasing}      \\
        $T_{\varphi}^i$                    & Pure Dephasing Time         & $\leq 2T_1$                &                                 \\
        \midrule
        $t_{\text{1-gate}}$            & 1-Qubit Gate Time         & $10^{-8}~s$      & \multirow{4}{*}{Depolarisation} \\
        $t_{\text{2-gate}}$           & 2-Qubit Gate Times         & $10^{-8}~s$      &                                 \\
        $\mathcal{F}_{\text{1-gate}}$  & 1-Qubit Fidelities         & $\lessapprox 1$  &                                 \\
        $\mathcal{F}_{\text{2-gate}}$ & 2-Qubit Fidelities         & $\approx 0.97$   &                                 \\
        \midrule
        $\varepsilon^\text{meas}_0$        & Error Rate of $\ket{0}$           & $\approx 1~\%$ &                                 \\
        $\varepsilon^\text{meas}_1$        & Error Rate of $\ket{1}$         & $\approx 1~\%$ &                                 \\
        \bottomrule
    \end{tabularx}
    \caption{Overview of the noise parameters describing an imperfect noisy QPU. Note that the values can significantly vary for different qubits within one QPU. The provided reference values are typical for the IBMQ devices used in this work.
    }\label{tab:noise_parameter}
\end{table}

\section{Methodology}
\label{sec:methodology}

\subsection{Parametrised simulator}

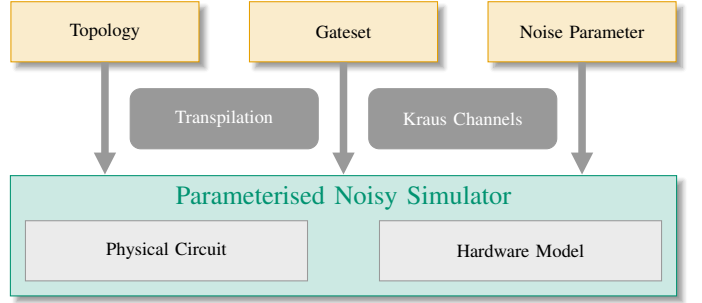
\begin{figure}
  % Base dimensions
\newlength{\WIDTH}\newlength{\HEIGHT}
\setlength{\WIDTH}{\columnwidth}
\setlength{\HEIGHT}{6cm}

\def\boxheight{0.8cm}
\def\boxwidth{2.5cm}
\def\boxdistx{0.5cm}
\def\boxdisty{1.5cm}

\tikzset{input/.style={rectangle, align=center, fill=lfd2!20, draw=lfd2, blur shadow, minimum height=\boxheight, minimum width=\boxwidth, font=\scriptsize, text=black}}
\tikzset{output/.style={rectangle, align=center, fill=lfd4!20, draw=lfd4, blur shadow, minimum height=2*\boxheight, minimum width=\WIDTH, font=\scriptsize, text=lfd4}}
\tikzset{interm/.style={rectangle, draw, align=center, fill=lfd3, draw=lfd3, rounded corners, minimum height=\boxheight, minimum width=\boxwidth, font=\scriptsize, text=white}}
\tikzset{arr/.style={-{Triangle[length=3mm, scale width=0.6]}, draw=lfd3, line width=1mm, font=\scriptsize}}
\tikzset{subbox/.style={rectangle, draw=lfd3, fill=lfd3!20, minimum height=\boxheight, minimum width=1.5*\boxwidth, font=\scriptsize}}
\tikzset{titletext/.style={text=lfd4, font=\normalsize}}

\begin{tikzpicture}

  \coordinate (tl) at (0, \HEIGHT);
  \coordinate (tm) at (0.5*\WIDTH, \HEIGHT);
  \coordinate (tr) at (\WIDTH, \HEIGHT);
  \coordinate (bl) at (0, 0);
  \coordinate (br) at (\WIDTH, 0);

  \node (topology) [input, anchor = north west] at (tl) {Topology};
  \node (gateset) [input, anchor = north] at (tm) {Gateset};
  \node (noise) [input, anchor = north east] at (tr) {Noise Parameter};
  \node (simulator) [output, below=\boxdisty of gateset.south, anchor=north] {};
  \node (subbox1) [subbox, anchor=west] at ($(simulator.west) + (2mm, -2mm)$) {Physical Circuit};
  \node (subbox2) [subbox, anchor=east] at ($(simulator.east) + (-2mm, -2mm)$) {Hardware Model};
  \node (simtext) [titletext] at ($(simulator.north) + (0, -3mm)$) {Parameterised Noisy Simulator};  
  \node (transpilation) [interm] at ($(topology.south)!0.5!(simulator.north)$) {Transpilation};
  \node (krauss) [interm] at ($(noise.south)!0.5!(simulator.north)$) {Kraus Channels};  
  \draw[arr] (noise.south) -- (simulator.north -| noise.south);
  \draw[arr] (gateset.south) -- (simulator.north -| gateset.south);
  \draw[arr] (topology.south) -- (simulator.north -| topology.south);

\end{tikzpicture}\vspace*{-1em}
  \caption{Summary of the noisy simulation stack, encapsulating topology, gate specifications, environment and transpiler.}
  \label{fig:parameterized_noise_model}
\end{figure}

The software components for the noisy simulation are shown in \autoref{fig:parameterized_noise_model}.
The three building blocks are the gateset, the topology (\ie the connectivity of the qubits), and the noise parameters of the gates and qubits.
The noise parameters, of which typical values are summarised in \autoref{tab:noise_parameter}, include the fidelities of the one- and two-qubit gates and the decoherence times of the qubits, as well as the readout error rate of the latter.

The performance of a quantum algorithm also strongly depends on efficient transpilation.
In this work we use the the Qiskit transpiler~\cite{qiskit} for translating circuits to the IBMQ device.
It is notable that the Qiskit compiler might introduce ancillae qubits, for instance to abbreviate swap-gates.
This may lead to an increased number of qubits in the transpiled circuit, posing potential challenges in the simulation on classical hardware, as well as in the evaluation of the results.
We therefore extend the Qiskit transpiler with our software stack, allowing to create fictitious QPUs, which can suit specific algorithms.
An ideal QPU architecture derived from this co-design approach could be fed back to hardware developers.

\subsection{Benchmark Circuits}
To evaluate the performance of the noise model, we analysed various problems formulated as quantum circuits.
The specific problem formulations are outlined as follows:
\subsubsection{Idle Circuits}
\label{sec:idle-method}

\begin{figure}[ht]
  \begin{subfigure}{0.38\columnwidth}
    \centering
    \begin{tikzpicture}[font=\small]
  \begin{yquant}
    qubit {$\ket{1}$} q[1];

    box {D [$t$]} q[0];
    measure q[0];
  \end{yquant}
\end{tikzpicture}
    \caption{$T_1$}
    \label{fig:idle_circ}
  \end{subfigure}
  \begin{subfigure}{0.58\columnwidth}
    \centering
    \begin{tikzpicture}[font=\small]
  \begin{yquant}
    qubit {$\ket{+}$} q[1];
    box {D [$\frac{t}{2}$]} q[0];
    x q[0];
    box {D [$\frac{t}{2}$]} q[0];
    h q[0];
    measure q[0];
  \end{yquant}
\end{tikzpicture}
    \vspace*{-1.5em}
    \caption{$T_2$ (Hahn echo experiment~\cite{Krantz_2019})}
    \label{fig:idle_echo}
  \end{subfigure}
  \caption{Quantum circuits for determining the $T_1$ and $T_2$ times. The qubits are kept idle for a time $t$ by applying a delay (D).}
\end{figure}
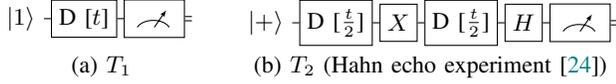

In order to verify amplitude damping and dephasing, and the corresponding decoherence times $T_1$ and $T_2$ in the noise data of IBMQ, we utilise dedicated \emph{idle circuits}.
This approach serves two primary objectives: firstly, to provide a practical sense of the magnitude of the individual noise effects; and secondly, to assess the accuracy of the noise parameter data.

\paragraph{$T_1$}
After initialising a qubit in the state $\ket{1}$ by applying a native $\hat{X}$-gate to the default initial state $\ket{0}$, a delay period $t$ is applied.
For this circuit, which is sketched in \autoref{fig:idle_circ}, eventually the qubit is measured in the $\{0, 1\}$ basis.

\paragraph{$T_2$}
To determine the effects of dephasing, we replicate the \emph{Hahn echo} experiment~\cite{Krantz_2019}, in which a qubit is prepared on the equator of the Bloch sphere, for instance by applying a Hadamard gate to the initial $\ket{0}$ state.
Subsequently, a delay $t$ is applied, similar to the previous experiment, but with a single $\hat{X}$-gate applied midway through the delay to account for inhomogeneous broadening mechanisms.
Finally, the state is measured in the $\{+, -\}$ basis, which is experimentally realised by applying another Hadamard gate before measurement.
The protocol for this experiment is sketched in \autoref{fig:idle_echo}.

\subsubsection{GHZ}
\label{sec:ghz}

\begin{figure}[ht]
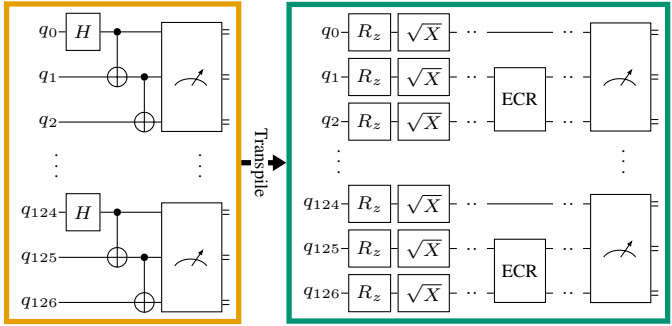

  \centering
  \begin{tikzpicture}
  \node[anchor=east, draw=lfd2, line width=2pt] (ghz_raw) at (0,0){\input{tikz-figs/ghz_raw}};
  \node[anchor=west, xshift=0.6cm, draw=lfd4, line width=2pt] (ghz_transpiled) at (ghz_raw.east){\input{tikz-figs/ghz_transpiled}};

  \draw[-{Triangle[scale=0.6]}, line width=2pt] (ghz_raw) -- node[pos=0.4, font=\scriptsize, rotate=-90, fill=white, inner sep=0] {Transpile} (ghz_transpiled);
\end{tikzpicture}
  \caption{Multiple three-qubit GHZ protocols, submitted in one circuit execution to an IBMQ system. In this simple case the transpilation step only consists of gate substitutions to the native gateset.}
  \label{fig:ghz}
\end{figure}
To determine the influence of two-qubit gate errors, we prepare the Greenberger-Horne-Zeilinger (GHZ)~\cite{greenberger90} state, which is a maximally entangled state.
For a general $n$-qubit system it is defined by
\begin{equation}
  \ket{\varphi_\text{GHZ}} = \frac{\ket{0}^{\otimes n} + \ket{1}^{\otimes n}}{\sqrt{2}}
\end{equation}
In particular, we focus on three-qubit GHZ circuits; to still gauge the performance of the full QPU, encompassing all qubits, we submit multiple GHZ circuits, acting on physically neighbouring qubits, in one execution to the IBMQ system.
Using Qiskit's~\cite{qiskit} transpilation level 0, only non-native gates are substituted with native gates, without circuit optimisation, which allows to conduct computations on selected qubits.
This process is shown in \autoref{fig:ghz}.

\subsubsection{Application-oriented benchmark quantum circuits}

To determine the impact of noise on practical quantum algorithms, we evaluate our noise model on the quantum approximate optimisation algorithm (QAOA), which is theoretically flexible in scalability and therefore applicable on current NISQ devices~\cite{bharti22}.

\paragraph{Quantum Approximate Optimisation Algorithm}
\label{sec:qaoa}

\begin{figure}[ht]
  \centering
  \begin{tikzpicture}[font=\small]
  \begin{yquant}
    qubit {$\ket{0}$} q[3];

    h q[0-2];
    box {$e^{-i\gamma_1 H_C}$} (q[0-2]);
    box {$R_x(\beta_1)$} q[0-2];
    box {$e^{-i\gamma_2 H_C}$} (q[0-2]);
    box {$R_x(\beta_2)$} q[0-2];
    measure q[0-2];

  \end{yquant}
\end{tikzpicture}
  \caption{QAOA circuit before the transpilation step for cost Hamiltonian $H_C$ and $p=2$.}
  \label{fig:qaoa}
\end{figure}
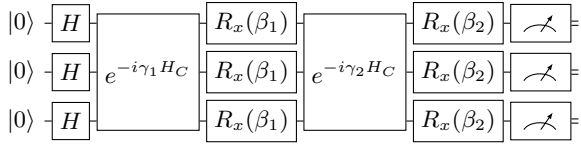

The QAOA employs a quantum circuit with $p \in \mathbb{N}$ layers of unitary operators defined by $2p$ parameters $\vec{\beta}, \vec{\gamma} \in \mathbb{R}^{p}$.
A QAOA layer $j$ consists of two unitaries:
\begin{equation}
  U_M(\beta_j) = e^{-i\beta_j\hat{H}_M},
\end{equation}
representing mixer Hamiltonian $\hat{H}_M$, and 
\begin{equation}
  U_C(\gamma_j) = e^{-i\gamma_j\hat{H}_C},
\end{equation}
based on the cost Hamiltonian $\hat{H}_C$, of which the ground state encodes the optimal solution to a given quadratic unconstrained binary optimisation (QUBO) problem (cf. \autoref{sec:problem}).
The mixer unitary $U_M$ typically consists of $\hat{X}$-rotations of size $\beta_j$ on each qubit, while the cost unitary $U_C$ uses single, or multi-qubit $\hat{Z}$-rotations of size $\gamma_j$.
The initial state of the QAOA algorithm is usually chosen as the ground state of $H_M$, in which each qubit is in an equal superposition of $\ket{0}$ and $\ket{1}$, prepared using a layer of Hadamard gates $H$.
A general example for a three-qubit QAOA circuit with $p=2$ is shown in \autoref{fig:qaoa}.

The repeated application of several QAOA layers corresponds to the discretised time evolution governed by the Hamiltonians $H_M$ and $H_C$.
It is known that the quality of the approximation increases with the number of layers~\cite{farhi14}, albeit the overall solution quality strongly depends on the parameter values $\vec{\beta}$ and $\vec{\gamma}$.
To find optimal parameters several approaches exist, which are inspired by adiabatic time-evolution~\cite{montanezbarrera24, zhou20}, or utilise classical optimisation routines~\cite{Powell1994, Periyasamy24}.
As in this work, we are mainly interested in the influence of noise and not in QAOA parameter optimisation, we employ the parameter initialisation strategy by Montanez-Barrera et al.~\cite{montanezbarrera24}, where they use fixed linear ramp (LR) schedules.
In the LR-QAOA protocol the parameters are given by
\begin{equation}
  \beta_j = \left(\frac{j-1}{p} - 1\right) \Delta_\beta, \qquad \gamma_j = \frac{j}{p} \Delta_\gamma
\end{equation}
for layer $j \in [1, p]$ and ramp sizes $\Delta_\beta, \Delta_\gamma \in [0,1]$.

\paragraph{Problem Description}
\label{sec:problem}

The QAOA is designed to optimise QUBO problems, to which any \textbf{NP} problem can be reduced~\cite{lucas14}.
The objective function is given by
\begin{align}
  E = \vec{x}^T Q \vec{x},
\end{align}
for a QUBO $Q \in \mathbb{R}^{n \times n}$, and $n$ binary variables $\vec{x}^T = (x_1, x_2, \dots, x_n) \in \{0,1\}^n$.
By switching from binary variables to spin variables ($\in \{-1,1\}$)~\cite{lucas14}, a QUBO can equivalently be described by an Ising model $\hat{H}_C$ that can be implemented in the QAOA.

In particular, we consider a problem from an industrially relevant use case, which is Job Shop Scheduling (JSS)~\cite{xiong22}.
The objective in JSS is to distribute a number of jobs $n_\text{j}$, whose production times $\{t_j\}$ vary, among a number of production machines $n_\text{m}$ in such a way that the variance of the total processing time per machine is minimised, and thus the machines are utilised optimally\footnote{We apply some simplifications to the original use case, such as regarding all machines equal and not considering job orders}.
In general, the number of different configurations increases exponentially with $n_\text{m}^{n_\text{j}}$.

In this work, we consider two encodings for the JSS problem:
(1) One-hot encoding, in which each variable (qubit) $x_{j,m}$ corresponds to the truth value of job $j$ being executed on machine $m$.
This may results in invalid schedules such as executing one job multiple times or not at all.
Accordingly, penalty terms are added to the objective function to suppress bit-strings corresponding to invalid states.
Furthermore, we use a (2) dense encoding, in which a machine on which the job $j$ is executed is represented in an integer register $|m_j\rangle$.
While the dense encoding requires less qubits, this comes at the cost of needing terms of higher order than quadratic.
Therefore, the problem may no longer be defined as a QUBO, but as a polynomial unconstrained binary optimisation (PUBO) problem.
However, higher-order terms are not an obstacle for a QAOA approaches, as several reductions from PUBO to QAOA circuits can be applied (see \eg Refs.~\cite{schmidbauer24_quick, schmidbauer24_qsw} for a review).
In this work, we use the implementation provided by the Qaptiva libraries~\cite{Qaptiva}.

The two resulting QAOA circuit classes have qualitatively different structures, summarised in \autoref{tab:overview_experiments}.
In this work, we transpile all QAOA circuits to the QPU by using Qiskit's transpiler with optimisation level 3, which, additionally to native gate substitutions, maps the circuit to suited qubits, and conducts further circuit optimisations~\cite{qiskit}.
The identical transpiled circuit (\ie with physical qubit mappings and native gates) is then executed on an IBMQ QPU and simulated on the Qaptiva.

\begin{table}[htbp]
    \centering
\begin{tabularx}{\linewidth}{@{}>{}l@{\hspace{.5em}}|XXXXXX}
\toprule
    \textbf{JSS Size (\# Jobs)} & 4 & 4 & 5 & 5 & 6 & 6 \\
    \textbf{Encoding} & Dense & 1-Hot & Dense & 1-Hot & Dense & 1-Hot \\
    \textbf{QAOA Depths} & $\leq 8$ & $\leq 24$ & $\leq 8$ & $\leq 16$ & $\leq 8$ & $\leq 16$ \\
    \textbf{\# Experiments} & 25 & 90 & 25 & 35 & 25 & 35 \\
    \midrule
    \textbf{\# Qubits} & 4 & 8 & 5 & 10-14 & 6 & 12-14 \\
    \textbf{\# 2-Qubit Gates}\(^{\text{*}}\) & 20 & 71 & 34 & 125 & 59 & 178 \\
    \textbf{Circuit Depth}\(^{\text{*}}\) & 62 & 114 & 97 & 168 & 138 & 168 \\
    \bottomrule
\end{tabularx}
    \caption{Overview of all LR-QAOA benchmark circuits
conducted on IBM heron device \texttt{ibm\_fez}. \# Experiments denotes the product of JSS instances and different QAOA
depths. The lower part of the table shows the number of qubits, the number of two-qubit gates and circuit depth for problem size (JSS Size). Numbers are averaged over stochastic transpiler passes. Rows marked with \(^{\text{*}}\) scale with QAOA depth and provide per-layer quantities.}
    \label{tab:overview_experiments}
\end{table}

\subsection{Measures of Distance}
\label{subsec:metrics}
A metric often used to compare probability distributions is the Hellinger distance (HD)
\begin{equation}
H\!D(p,q) =  \sqrt{ \frac{1}{2}\sum_i (\sqrt{p_i}-\sqrt{q_i})^2}
\end{equation}
which is closely related to the fidelity
\begin{equation}
F(p,q) = \sum \sqrt{p_i}\sqrt{q_i} 
\end{equation}
via
\begin{equation}
H\!D(p,q)^2 = 1 - \sum_i \sqrt{p_i}\sqrt{q_i} = 1 - F(p,q).
\end{equation}

Following Ref.~\cite{Georgopoulos_2021},  we use HD to compare probability distributions obtained by running GHZ circuits on the IBMQ device and on the Qaptiva simulator (cf.~\autoref{sec:exp_results}). Note that to apply this metric,
probabilities $p_i$ for every state $i$ of the computational basis must be estimated from relative counts, which gets infeasible with an increasing qubit count as it requires exponentially increasing number of shots. 
The same problem arises for other commonly used distance measures such as total variation distance, Jensen-Shannon divergence or Kullback–Leibler divergence.
We therefore propose an alternative way to quantify the difference between QPU results and simulation: It should be applicable to larger qubit numbers, and should ideally be related to the solution quality when applied to QAOA circuits.

\subsection{Maximum Likelihood Estimation of Best Solution Probability}
Based on the experimental observation that probability distributions resulting from an LR-QAOA circuit~\cite{montanezbarrera24} can be described well by an exponential function in the state energy, we propose to characterise these distributions
as
\begin{equation}\label{eq:boltzmann}
p(x) = Z^{-1} \exp(-\zeta E(x))    
\end{equation}
where $x$ labels the states of the computational basis, $\zeta$ is a free parameter, and $Z := Z(\zeta)$ is a normalisation factor to ascertain
\(\sum_x p(x) = 1\).
It holds that
\begin{equation}
Z(\zeta) = \sum_x{ \exp(-\zeta E(x))}.
\label{eq:normalization}
\end{equation}
Note the formal similarity with the Boltzmann distribution in statistical physics where $\zeta$ signifies an inverse temperature.

We expect that the exponential energy dependency of the probabilities is directly related to the success of QAOA or similar methods, as the density of states at the energy minimum (and also at the energy maximum) is typically exponentially smaller than at medium energies.
To find a state near the energy minimum with a probability that is significantly larger than the uniform probability $2^n$, the probability distribution must drop exponentially when moving from the energy minimum towards medium energies.

However, our approach is not limited to an exponential form, but can accommodate any piecewise analytical function. Notably, this includes probability distributions that can  generally be represented by expanding the exponent (related to their moment). Incorporating higher-order terms in the exponent of \autoref{eq:boltzmann} allows us to describe a broad class of distributions. While physical intuition on the fitting parameters might be limited in such cases, it does not reduce applicability of the current method, and is well suited to overcome the problem of distribution suffering from low statistics.

\subsubsection*{Maximum Likelihood Estimation of $\zeta$ and $\tilde p$}
Assuming he evaluation of an LR-QAOA circuit with $N$ shots that results in state $x_i$ for $n_i$ times ($\sum_i n_i = N$), we aim to determine the value of $\zeta$ such that this result is matched sufficiently accurate.
Under the assumption that the probability to find some specific state $x$ is given by \autoref{eq:boltzmann},
the joint probability to find the given result is
\begin{equation}
P(\zeta) = N!\prod_i \frac{p(x_i;\zeta)^{n_i}}{n_i!}.
\label{eq:multinomial_likelihood}
\end{equation}

Now, the maximum likelihood estimator of $\zeta$ is the value of $\zeta$ that maximises $P$. It is equivalent but computationally more convenient to maximise $\log(P)$.
Ignoring terms that do not depend on $\zeta$ results in the condition

\begin{equation}
  \begin{aligned}
  0&=\sum_i n_i \frac{d}{d\zeta}\log(p(x_i;\zeta)) %\\
  = - N \frac{d}{d\zeta} \log(Z)- \sum_i n_i E(x_i) \\
  &= N\left( \langle E \rangle_\zeta - \frac{1}{N}\sum_i n_iE(x_i)\right),
  \end{aligned}
\end{equation}

where we used
\begin{equation}
\langle E \rangle_\zeta := Z^{-1}\sum_i{ E(x_i)\exp(-\zeta E(x_i))} = -\frac{d}{d\zeta} \log(Z).
\end{equation}

Finally, the condition for $\zeta$ reads
\begin{equation}
\label{eq:expect_e_cond}
\langle E \rangle_\zeta = \frac{1}{N}\sum_i n_iE(x_i).
\end{equation}

The calculation of $\langle E \rangle_\zeta$ involves a sum over the whole state space.
Nevertheless, and in contrast to the HD, it can still be calculated for large amounts of qubits: Including, in particular, cases in which the required number shots to estimate the whole probability distribution is beyond practical limits.

The statistical uncertainty of the fitting approach is evaluated by varying the fitting parameter from the optimal value such that the resulting likelihood is only $60\%$ of the maximum likelihood. This parameter value corresponds to one standard deviation.

\subsubsection*{Taking Noise Effects into Account}

Noise effects, especially measurement errors, lead to non-vanishing probabilities of high energy states. We account for this by adapting the probability distribution in \autoref{eq:boltzmann} by
\begin{equation}\label{eq:boltzmann_with_noise}
p(x) = Z^{-1} \left( \exp(-\zeta E(x)) + \delta \right)   
\end{equation}
with an additional parameter $\delta$. While this expression for $p(x)$ lacks a simple analytic relation similar to \autoref{eq:expect_e_cond}, $P(\zeta,\delta)$ can still be maximised numerically. We maximise using the \verb|scipy.optimize| toolkit.

In \autoref{sec:exp_results}, we use $\tilde p := p(x^*)$ with $x^*$ being a lowest energy state and $p(x)$ given by \autoref{eq:boltzmann_with_noise} with optimised parameters $\zeta$ and $\delta$ to characterise the probability distributions resulting from QAOA circuits. 
$\tilde p$ is closely related to the probability $p_0$ of finding an optimal solution, which is an intuitive and practically relevant parameter.
Assuming (a) the probability distribution \autoref{eq:boltzmann_with_noise} describes the circuit outcome properly, and (b) the lowest energy state is $n$-fold degenerated, we have $p_0 = n\tilde p$.

\section{Noise Model Validation Results}
\label{sec:exp_results}
\subsection{Idle Noise Circuits}
\begin{figure}[]
    \centering
    \includegraphics{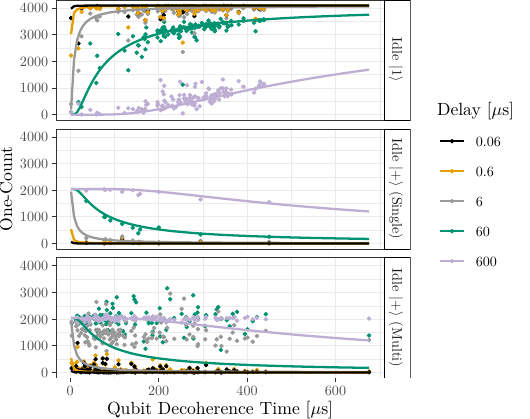}
    \caption{Experimental results for the decay of single qubit states during the $T_1$ and $T_2$ experiments using $N = 4096$ shots and various delay times $t$. The different delay times are distinguished by colour. Each dot corresponds to one qubit and shows how often this qubit has been found in the 1-state. The x-axis displays the decoherence time (\(T_{1}\) for the top panel, \(T_{2}\) for the other two panels) as reported by IBMQ for each qubit. Solid lines show the behaviour as expected from model \autoref{eq:dephasing}. The top panel shows the measured decay for decoherence time $T_1$, compared with the theoretical exponential decay model $N e^{-t/T_1}$ for each qubit.
    The middle and bottom panel display the dephasing behaviour from the Hahn echo experiment compared to the theoretical model $\frac{N}{2}(1 - e^{-t/T_2})$.
    The middle panel refers to individually conducted qubit experiments, the bottom to simultaneously evaluated qubits.}
    \label{fig:t_exp}
\end{figure}

To evaluate the relaxation time $T_1$ and $T_2$, provided in the noise model data by IBMQ~\cite{ibm}, we submit the idle circuits, as described in \autoref{sec:idle-method} on the 127 qubit device \texttt{ibm\_kyiv}.
As the $T_{1}$ parameters of the individual qubits in the system are different, the decay of the prepared states can be quantitatively validated by setting the delay parameter t to different selected values (see colour coding in \autoref{fig:t_exp}). According to the hypothetical amplitude damping of the excited state (see \autoref{eq:dephasing}), the number of measurements with result \enquote{1} for qubit $i$ if repeating the experiment $N$ times, is given by $N_i = N e^{-t/T_{1,i}}$ where $T_{1,i}$ is the qubit specific $T_1$-relaxation time.

In the top of \autoref{fig:t_exp}, this function is shown as solid lines for 4 different values of $t$. Each dot corresponds to one qubit. Its y-coordinate is the measured 1-count and its x-coordinate is the latest available $T_1$ calibration data for that qubit as reported by IBMQ. The experimental results for $T_1$ align with the values provided by IBMQ, when submitting the circuit from \autoref{fig:idle_circ} simultaneously on all 127 qubits.
The measurement outcomes are accumulated over $N=4096$ preparations of the same circuit.

When applying the Hahn echo circuit from \autoref{fig:idle_echo} for $T_2$ similarly on all qubits simultaneously, we obtain the results shown at the bottom of \autoref{fig:t_exp}.
It is evident that results do not follow the hypothetical dephasing curves.
Potential explanations for this deviation may include unintended interactions between the qubits, known as \emph{cross-talk}~\cite{murali20}, or a perturbation of the echo mechanism caused by temporal shifts of the operations and measurements.

When, instead of evaluating the circuit on all qubits simultaneously, we conduct separate evaluations of 10 individual qubits, with varying reported $T_2$ times, for a total of $N=4096$ shots of the circuit the experimentally measured number of \enquote{1} results aligns with the theoretical dephasing result of $\frac{N}{2}(1-e^{-t/T_2})$
as shown in the middle facet of \autoref{fig:t_exp}. 

\subsection{GHZ}

\begin{figure}[]
  \begin{subfigure}{\linewidth}
    \centering
    \includegraphics{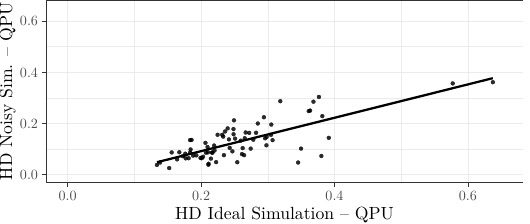}
    \caption{%
      Ideal and noise model comparison.
    }
    \label{fig:hellinger}
  \end{subfigure}
  
  \begin{subfigure}{\linewidth}
    \centering
    \includegraphics{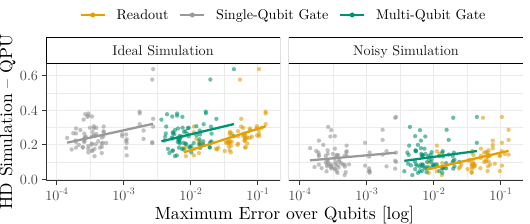}
    \caption{%
      Dependency on the types of errors.
    }
    \label{fig:error_dependency}
  \end{subfigure}
  
  \begin{subfigure}{\linewidth}
    \centering
    \includegraphics{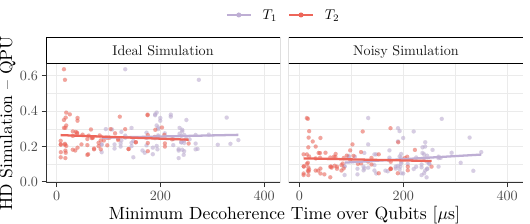}
    \caption{%
      Dependency on the decoherence times.
    }
    \label{fig:ghz_t12_dependency}
  \end{subfigure}
  \caption{%
    Hellinger Distance (HD) of the GHZ experiments.
    Each point refers to the evaluation of one three-qubit GHZ circuit, simultaneously evaluated on the \texttt{ibm\_osaka} QPU, as introduced in \autoref{sec:ghz}.
    Each colour comprises the full set of GHZ circuits, but in dependency of different properties of the QPU.
    Lines represent a linear trend.
    In (a), the x-axis describes the HD for the evaluation on QPU and an ideal simulation.
    The noise data of the same set of qubits is used for a simulation with our noise model, of which the HD to the QPU results is shown on the y-axis.
    Subfigures (b) and (c) show the dependency on the respective maximum error and minimum decoherence time over the corresponding qubits, which is shown on the x-axis.
  }
  \label{fig:ghz_results}
\end{figure}

The results presented in \autoref{fig:ghz_results} show the differences between the IBMQ device \texttt{ibm\_osaka}, and our simulations for the GHZ experiments described in \autoref{sec:ghz}.
As illustrated through \autoref{fig:hellinger}, there exists a positive correlation between the HD for ideal and noisy simulations, indicating that the simulation is less accurate if the overall errors are large.
Notably, the quality of the simulation depends on the accuracy of the noise model as well as on the accuracy of the available noise parameters.

As GHZ circuits are sensitive to single-qubit failures, \autoref{fig:error_dependency} shows the dependency of the HD on the maximum occuring errors throughout the three qubits of the GHZ circuit.
In that the single-qubit error refers to $X$ and $\sqrt{X}$ gates; the multi-qubit gate error refers to the ECR-gate error.
Similarly, \autoref{fig:ghz_t12_dependency} shows the dependency on the minimum $T_1$ and $T_2$ decoherence times.
The left panels show the HD of results from ideal (noiseless) simulation and from \texttt{ibm\_osaka} QPU as a function of the strength of the different noise channels as described by the IBMQ noise data.
The slope of the linear trend lines in these panels gives a measure of how much the quality of the noisy modelling depends on the different noise errors.
We see that the readout and gate errors are the primary sources of errors for the GHZ circuits considered here while the decoherence times have a smaller impact.
The right panels show the HD of our noisy simulations and results from \texttt{ibm\_osaka} QPU.
The smaller slope observed in the right panel of \autoref{fig:error_dependency} for the readout and gate error suggests that these can adequately described with our noise model, albeit tend to be modelled less accurate.

The HD between noisy simulation and QPU is on average $0.13 \pm 0.07$ for a three-qubit circuit, which is comparable to the HD values obtained by the noise models in Ref.~\cite{Georgopoulos_2021} for the same circuit size.
This indicates that our stack for noise-model performs similar to other state-of-the-art tunable noise models.
Two outliers in the execution on the real device are identified (\ie, points where HD between ideal simulation and QPU exceeds 0.5), suggesting that some qubits are not functioning properly.
Consequently, these outliers also provide inaccurate noise data.

We iterated the analysis using other well known distance measures (total variation distance, Jensen-Shannon divergence, Kullback–Leibler divergence) instead of HD, arriving at identical conclusions as the overall trend is equal for all tested metrics (distance ranges differ).

\subsection{Noise Model Validation in a use case circuit (QAOA)}
\label{sec:validation}

In this section we utilise the metric $\tilde p$ as introduced in \autoref{subsec:metrics} for characterising the probability distributions obtained from running the circuits on the QPU or simulating them. As for small circuits we observed that the effect of amplitude damping and pure dephasing only contributes insignificantly to the overall error (cf.~\autoref{fig:ghz_t12_dependency}), we neglect the impact of these types of noise in the simulations of QAOA.

\begin{figure}
    \centering
    \includegraphics[]{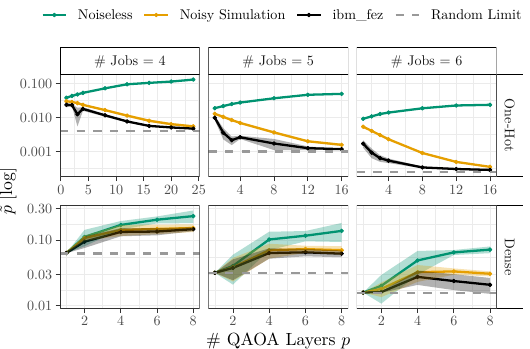}
    \caption{The behaviour of $\tilde{p}$ as obtained from a maximum likelihood estimation, as a function of the QAOA depth. The black line corresponds to the value obtained from the experiment conducted on the IBM Heron processor \texttt{ibm\_fez}. The yellow line corresponds to the simulation performed with the Qaptiva and the noise model described above, using the noise parameters that correspond to the calibration data at the time of the experiment. The teal line refers to an ideal simulation without noise.}
    \label{fig:px_vs_qaoa_depth}
\end{figure}

To quantify the accuracy of the simulation we compare $\tilde{p}$ as derived from the QPU results with those derived from our noisy and noiseless simulations. 
\autoref{fig:px_vs_qaoa_depth} shows the values of $\tilde p$ averaged over the instances of the JSS problem, summarised in \autoref{tab:overview_experiments}, in dependency of the QAOA depth. 
The dependency of the solution quality on the QAOA depth is an important information for practical considerations of solving a given optimisation problem. 
As to be expected, the noiseless simulation suggests that an increase in QAOA-depth improves $\tilde{p}$, our measure for the success rate of the overall circuit. Taking noise into account, either by the noise model or by a real QPU one can conclude a maximal useful QAOA depth. 
\autoref{fig:px_vs_qaoa_depth} shows that the trend of the curves can qualitatively be extracted from the simulated data. For circuits with a low number of qubits the deviation of $\tilde p$ between noisy simulation and QPU results is lower than 10\%. This deviation increases for larger numbers of jobs but the order of magnitude and the above conclusions remains the same.

A problem-independent measure of QAOA performance is the difference between the probability to find the best solution using QAOA and the probability to find the best solution with a uniformly random bit-string, which is $p_0=1/2^n$.
The gain ratio 
\begin{align*}
    r = \frac{\tilde p_\text{noisy} - p_0}{\tilde p_\text{noiseless} - p_0}
\end{align*}
quantifies the impact of noise on this potential gain.
$\tilde p_\text{noisy}$ stands for either $\tilde p$ as found from circuit executions on the real QPU or as found from noisy simulations. 

\begin{figure}
    \centering
    \includegraphics[]{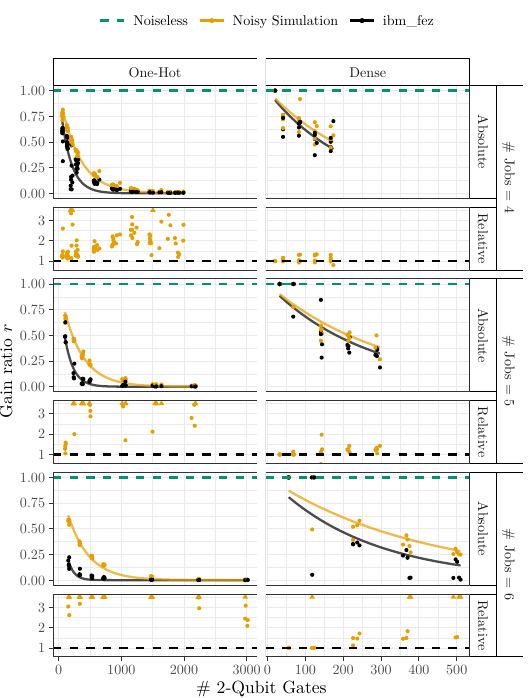}
    \caption{The gain ratio $r$ (\ie the advantage of the quantum algorithm that is left when conducted on a noisy device) as a function of the number of two-qubit entangling gates (\ie CZ) for real QPU (black) and simulation (yellow). 
    Each point represents one conducted experiment or its simulative prediction respectively.
    The lines represent an exponential fit of the data points.
    It starts at $1$ for small circuits and decreases to $0$ for larger circuits ($\approx 1000$ CZ gates). The smaller panels show the individual relative error (\ie the ratio of gain ratios) between each prediction with its real QPU counterpart.  }
    \label{fig:delta_p}
\end{figure}

A value of $1$ for the gain ratio is achieved in the limit of a noiseless execution and becomes $0$ if there is no gain of the algorithm with respect to a random choice.
In \autoref{fig:delta_p} this quantity is shown for $\tilde p_\text{noisy}$ as determined from QPU and the corresponding noisy simulations. 
Independent of the benchmark circuit, the gain ratio loosely follows an exponentially deteriorating pattern with increasing 2-qubit gate count. This is be observed in both, simulation and experiment. 
The half-value interval of the gain ratio is $\approx 220$ 2-qubit gates throughout all simulated results.
On real hardware this interval is smaller with $\approx 180$  2-qubit gates for healthy QPU executions, indicating and quantifying the systematic underestimation of the effect of noise.
The effect of statistical fluctuations of the value $r$ due to the finite number of shots and the fitting procedure has been evaluated and found to be small ($\Delta r\approx 0.02$) compared to the spread between different circuit instances. The cause for this spread is likely the different qubit location on the QPU at which the circuit was executed on, yielding different noise parameter. 
Systematic uncertainties due to the inaccuracy in the calibration data have not yet been considered and are subject to future studies.

A benchmark-inclusive analysis demonstrates a strong overall agreement between the model and the actual QPU: Disregarding the strongly deviating outliers, the possible reasons for which are discussed below, a linear fit was applied to the data points of the relative deviation (lower panels of~\autoref{fig:delta_p}). This results in a deviation of less than 20\% in the limit of small circuits. This degree of mis-modelling increases with the size of the circuits in the order of 100\% per $\approx0.1~$ms circuit execution time or $\approx500$ applied 2-qubit gates.
Nevertheless, this prediction may serve as an upper limit for certain scalability considerations, as discussed in~\autoref{sec:scaled_fid}.

When examining the individual benchmark scenarios separately, the accuracy of the noise model shows significant variability.
While the model accurately describes the 4-job case in one-hot encoding, it is overly optimistic for the 5- and 6-job cases.
A detailed analysis of the experimental state distributions for these circuits indicates that the most frequent states contain fewer ones than number of jobs.
However, a distinctive feature of the one-hot encoding approach is that the valid (\ie, the lowest-energy) states contain as many ones as there are jobs in the JSS problem.
A direct correlation can be established between each probable state and one of the valid states, which only differ by a Hamming distance of 1.
This suggests that the model fails to describe certain aspects that explain the transition from ones to zeros.
In fact, since the model was developed in the context of smaller circuits, qubit relaxation was considered negligible and was not simulated. The missing modelling of thermal relaxation into the ground state naturally over-predicts ones and thus states with ones equal to the number of jobs (\ie, valid states), leading to a far higher expectation for $\tilde{p}$ than observed. This explanation aligns with the linearly increasing systematic deviation for the overestimated benchmarks.
Another possible explanation is imprecise calibration data. A \texttt{prep1\_meas0} value (\ie the probability of preparing $\ket{1}$, but measuring zero) of 10\%-15\% for some of the used qubits may also account for the reduced performance observed on the QPU.
Such a value is not implausible, as errors with similar orders of magnitude were determined for other qubits on the same QPU.
Furthermore, this explanation is consistent with the following observation made in the 4-job one-hot case:
circuits with a QAOA depth of 3 were submitted to IBM on a different date than the circuits with QAOA depth of 2 and 4.
The observed performance at this point in time, in terms of $\tilde{p}$, was significantly worse, despite no fundamental differences in the circuits themselves.
A similar observation can be made in the 6-jobs dense-encoded scenario, where there are clearly two distinct clusters.

\section{Application: Scaled Fidelity}
\label{sec:scaled_fid}
The parametrised noise model offers the possibility of scaling the fidelity to gauge the usability of solution approaches on future, improved QPUs.
For this purpose, the error rate is reduced in steps of half an order of magnitude.
The simulation is validated against the experimental QPU noise data, which corresponds to the scaling factor 1.
In contrast to \autoref{sec:validation}, the utilised noise parameters are constant across all qubits, corresponding to the median value of the respective parameter of the \verb|ibm_kyiv|, to represent a more general contemporary IBM Eagle processor.

\begin{figure}[hb]
    \centering
    \includegraphics{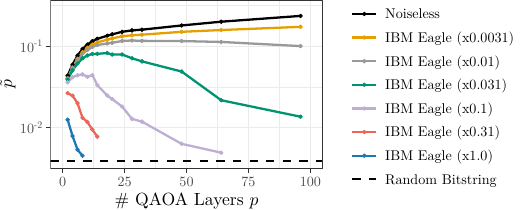}
    \caption{The behaviour of $\tilde p$ as obtained from a maximum likelihood estimation, as a function of the QAOA depth. The different lines correspond to different gate error rates, with the blue line corresponding to the median error rate of a current IBM Eagle processor.}
    \label{fig:scaled_fid_4on2}
\end{figure}

\autoref{fig:scaled_fid_4on2} shows the quality metric $\tilde{p}$, under the influence of the improved QPUs for an exemplary 4-job JSS instance, dependent on QAOA circuit depth.
It is apparent that the optimal choice of QAOA depth improves progressively with more fault-tolerant systems.

\section{Discussion}
\label{sec:disc}

We tested the validity of our noise model with conceptual circuits such as the Hahn echo experiment and simultaneous three-qubit GHZ experiments.
We found that in most tests the simulated results were in agreement with the experimental results, with small corresponding HDs between $0.03$ and $0.12$.
However, in some cases, the simulation deviates considerably from the experiment, indicating that the noise parameters provided by IBM are not always accurate.
Simultaneous Hahn echo experiments on a QPU show a significant difference in the final state compared to the expectation due to dephasing.
The dephasing seems to occur faster with simultaneous operations, which we suspect is a consequence of cross-talk~\cite{murali20}, that is not captured by our noise model.

Furthermore, practical LR-QAOA circuits were tested, which were generated based on a real-world JSS problem.
The benchmark circuits ranged from problem sizes using 4 up to 12 qubits, some of which are even executed on 16 physical qubits after transpilation.
We presented a method to assess the quality of probability distributions in large state spaces, incorporating a maximum-likelihood fit.
This method was applied to validate the QAOA benchmark circuits.
The likelihood fit allows us to characterise probability distributions over large state spaces ($2^n$)
where the full distribution is not accessible due to a lack of statistics ($4000$ shots).
In particular, in the case of the LR-QAOA, the probability of the best solution, $\tilde{p}$, can be derived from the fit, even if this probability is small.
The parameter $\tilde{p}$ allows to validate the noise model by comparing the resulting values of experiment and simulation.
The values for $\tilde{p}$ generated by the simulation are in good agreement with what is derived from the experiment, with a deviation of $\lessapprox  10\%$ for short and medium-sized circuits. Still, the order of magnitude is well estimated by the applied noise model for larger circuits.
The deviation here is systematic, in that the simulation delivers a $\tilde{p}$ that is slightly too optimistic.
Thus, by applying our method we are able to quantify the degree of precision of the examined noise models, and are able to identify the use cases, where the model is not sufficiently accurate.
Furthermore, we occasionally experienced IBMQ QPUs delivering unusually bad results:
Almost identical logical circuits deliver a significant large probability of success, which only 12 hours later yield significantly worse results.
The noisy simulation is not exposed to the fluctuations described here, suggesting that it can offer a more stable environment for the development of algorithms and subroutines.

We attempted validations with different QPU architectures. Unfortunately, many vendors conduct hidden error mitigation and use internally (and intransparently) transpiled circuits, which leads to unsound results, as evaluating identical circuits for the real hardware and the simulation is crucial to validate the noise model.
We need to leave this issue to future work that can be conducted once vendors provide appropriate direct system access.

In addition, the parametrisable noise model was employed to illustrate that the exemplary LR-QAOA approaches necessitate error rates that are orders of magnitude smaller to generate a benefit.
Nevertheless, this suggests the existence of promising algorithmic pathways for more fault-tolerant systems.

\section{Conclusion and Outlook}
\label{sec:concl}

In this work, we presented a modular parametrisable noise model, in which core QPU properties, such as topology, gateset and noise parameters can be individually configured.
This makes it a useful and (mostly) reliable tool to evaluate the performance of current and future quantum software.
Scaling up simulations to larger problem sizes, and also other types of problems in future work will contribute to the practicality and accuracy of current noise models.
Furthermore, our custom application-centred noise model quality metric enables to investigate the performance across different types of QPUs, albeit we need to leave a concrete instantiation to future work.

Generally, noise modelling in quantum computing is a dynamic and evolving field with significant implications for software engineering.
By addressing challenges related to noise data precision, and the development of algorithm-specific metrics, we can enhance the accuracy and practicality of noise models.
These advancements will not only improve the reliability of quantum algorithms but also pave the way for more efficient and effective quantum software in the future.

\appendix

\subsection{Source Code Availability}
A package with QUARK~\cite{rudi22} modules providing bindings to the Qaptiva~\cite{Qaptiva} libraries and containing our stack for the parametrisable noise model, is currently under active development and planned to be published in QUARK.
The experimental data, as well as code for data evaluation and plot generation is available at \repro.

\noindent\textbf{Acknowledgements}
The authors acknowledge support by \blackout{the German Federal Ministry of Education and Research (BMBF)}, funding program \blackout{\enquote{Quantum Technologies—--From Basic Research to Market}}, grant numbers \blackout{13N16092} and \blackout{13N16094}. \blackout{WM} acknowledges support by the \blackout{High-Tech Agenda of the Free State of Bavaria}.
We thank \blackout{Christopher Sowinski} for his support in accessing the quantum systems and acknowledge the use of IBM Quantum services for this work. The views expressed are those of the authors, and do not reflect the official policy or position of IBM or the IBM Quantum team. 

\printbibliography
\end{document}